# Self-biased SAW Magnetic Field Sensors Based on Angle Dependent Magneto-acoustic Coupling

Wenbin Hu, Mingxian Huang, Huaiwu Zhang, Feiming Bai*

State Key Laboratory of Electronic Thin Films and Integrated Devices, University of Electronic Science and Technology, Chengdu, 610054, China

* To whom correspondence should be addressed. Electronic mails:

fmbai@uestc.edu.cn

ABSTRACT

Surface-acoustic-wave (SAW) based devices have emerged as a promising technology in magnetic field sensing by integrating a magnetostrictive layer with the giant $ΔE/ΔG$ effect. However, almost all SAW magnetic field sensors require a bias field to obtain high sensitivity. In addition, the true nature of magneto-acoustic coupling still presents a major challenge in understanding and designing of this kind of devices. In current work, a dynamic magnetoelastic model for the $ΔE/ΔG$ effect is established in consideration of the important role of the dipole-dipole interaction. The model is also implemented into a FEM software to calculate the resonance frequency responses of multiple fabricated sensors with different $ψ$ angles between of the acoustic wave vector and the induced uniaxial magnetic anisotropy. The measured results are in excellent agreement with the simulated ones. A strong resonance frequency sensitivity (RFS) of 630.4 kHz/Oe was achieved at zero bias field for the device with optimized $ψ$ angle. Furthermore, the $RFS$ measurements along different directions verify its vector-sensing capability.

**I. Introduction**

Magnetic field sensing is critical for many applications including positioning, navigation, electrical current monitoring and biomagnetic field detection, etc. Recently, the $\Delta E/\Delta G$ effect has been used to detune cantilever [1–3], surface acoustic wave (SAW) [4–15] or bulk acoustic wave (BAW) [16–19] devices coated with a magnetostrictive film. The $\Delta E/\Delta G$ effect is known as the modification of elastic modulus with respect to a magnetic field. Giant $\Delta E$ effect exists in amorphous magnetostrictive materials with high saturation magnetostriction and low anisotropy [20]. Taking SAW devices for example, the phase velocity of the piezoelectric substrate becomes dispersive upon depositing a magnetic layer onto [4–8] or in-between [9–13] the interdigital electrodes (IDTs), or simply replacing the nonmagnetic IDTs with magnetic ones [14,15]. An ultrahigh DC magnetic field sensitivity of 2.8 Hz/nT and a limit of detection of 800 pT were reported by Li et al. using a AlN/FeGaB resonator [17]. Additionally, a very low magnetic noise level of 100 pT/ Hz and a bandwidth of 50 kHz have been demonstrated by Kittmann et al. [11] using a Love-mode delay line. Even higher sensitivity has been achieved by Schmalz et al. [13] using a Love-mode SAW magnetic-field sensor upon optimizing the IDT pitch width. This is because the Love wave, also called horizontal shear wave, propagates parallel to the magnetic film plane and are thus sensitive to changes of shear modulus $G$, which is more pronounced than the $\Delta E$ effect under a magnetic field.

However, these attractive sensitivity values were realized by applying certain bias magnetic fields (4 Oe for FeCoSiB, 12 Oe for FeGaB and even 400 Oe Terfenol-D) [11,17,21]. When used for navigation or biomagnetic applications, a precision range of only a few Gausses or even lower is sufficient. A large bias field inevitably makes a SAW sensor bulky, and consumes more power. Although self-biased magnetic sensors were previously reported by Liu et al. [5], the non-zero resonance frequency sensitivity (*RFS*) was attributed to the relatively high residual magnetic moment and large coercive field, contributing to the non-zero change of the Young's modulus at zero magnetic field. However, a large magnetic hysteresis causes poor sensing linearity. In addition, a true vector sensor characterizes with the highly selective sensitivity to one direction but not to the other vertical directions, which is absent in Ref. [5]. Alternatively, the exchange bias effect can be employed to design self-biased sensors [3,22]. However, an antiferromagnetic layer can only effectively pin a neighboring ferromagnetic layer below a certain thickness (typically <100 nm), which clearly set a limit on the magneto-acoustic coupling and thus the sensitivity of the magnetic-field sensors.

In current work, we propose a novel self-biasing approach based on a dynamic magnetoelastic model of the $\Delta E/\Delta G$ effect. Vector SAW magnetic-field sensors were designed and fabricated by adjusting the angle between the in-plane induced uniaxial anisotropy and the SAW propagation direction. A very high *RFS* of 630.4 kHz/Oe was demonstrated at zero bias field.

This paper is structured as follows. In Section II, we provide a dynamic magnetoelastic model for the $\Delta E/\Delta G$ effect (Section. II A) and a magneto-acoustic FEM model of a multi-layered SAW structure (Section. II B) to calculate the resonance frequency response of SAW magnetic sensors. Then the methods of device fabrication and measurement are presented in Section III. Section IV is devoted to the discussion of experimental results and the impact of dipole-dipole interaction on the magneto-acoustic coupling (Section. IV A), which was used to design self-biased vector magnetic field sensors. (Section. IV B).

## II. THEORY

Fig. 1 schematically shows the configuration of the SAW magnetic field sensor. A ST-cut 90°X quartz is selected as the piezoelectric substrate. The Love wave acoustic mode is excited by the IDT deposited on the piezoelectric crystal and resonates from reflections of the short-circuited reflector on both sides. A ferromagnetic FeCoSiB film is deposited over the entire IDTs and reflector grating, with a $SiO_2$ waveguide layer.

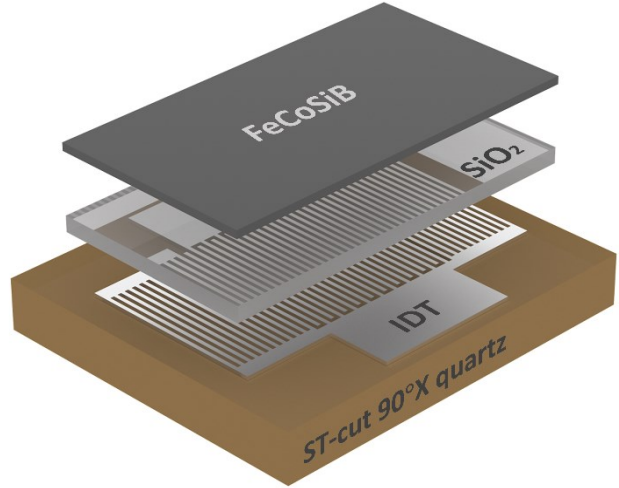

FIG. 1. Schematic illustration of multilayered Love-mode SAW resonator.

As an important parameter for evaluating device performance, the magnetic field sensitivity can be described by the resonance frequency sensitivity (*RFS*) for a magnetic SAW resonator,

$$RFS = \frac{\partial f}{\partial H} = \frac{\partial f}{\partial v} \cdot \frac{\partial v}{\partial c} \cdot \frac{\partial c}{\partial H} \qquad (1)$$

where *f*, *v* and *c* are the resonance frequency, the phase velocity and the elastic modulus of the resonator, respectively.

As can be seen, *RFS* is determined by three factors. The foremost factor $\frac{\partial c}{\partial H}$ represents the variation of elastic modulus with an external magnetic field, which is more familiar as the *ΔE/ΔG* effect. $\frac{\partial c}{\partial H}$ arises from the magnetostrictive strain, and the static ΔE/ΔG effect has been modeled in previous works [23–27].

Upon applying an external magnetic field, the phase velocity of SAW changes due to the variation of the effective elastic modulus of the magnetostrictive/piezoelectric multilayer. The $\frac{\partial v}{\partial c}$ term depends on the ratio between the magnetostrictive layer thickness and the acoustic wavelength, the acoustic impedance of materials and the acoustic mode of the structure. In current multi-layered resonator, the Love wave propagates unevenly in each layer, making it difficult to determine $\frac{\partial v}{\partial c}$ by analytical expressions, particularly when coupled with micro-magnetics, which increases the complexity of the system. Therefore, a magneto-acoustic coupling model is built in Sec. IIB, using the finite element method.

Finally, the $\frac{\partial f}{\partial v}$ term is determined by the geometric period of the interdigital electrodes (IDTs). Although magnetostriction also affects the wavelength and causes the resonance frequency shift, this effect is negligible [28]. So, $\frac{\partial f}{\partial v}$ is equal to $1/\lambda$ in the followed calculation.

**A. The dynamic magnetoelastic model of ΔE/ΔG effect**

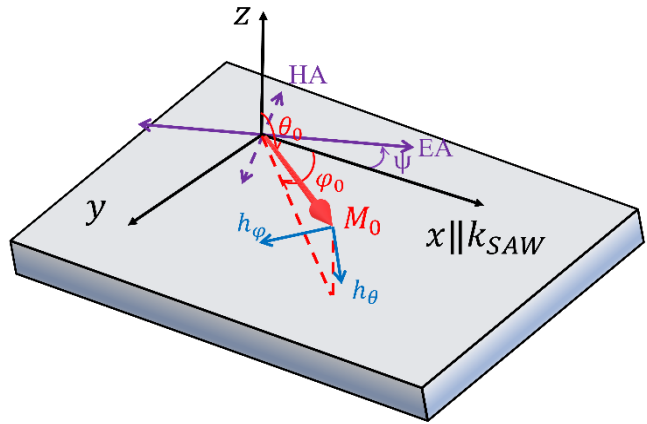

FIG. 2. Schematic illustration of the Cartesian coordinates used in the calculation.

As shown in Figure 2, a Cartesian coordinate system is built to model the magneto-elastic wave in the ferromagnetic film with *x*-axis parallel to the acoustic wave vector, $\boldsymbol{k_{SAW}}$. Let's consider a spontaneously magnetized ferromagnet

$$\boldsymbol{M} = \boldsymbol{m}M_s, \text{ with } |\boldsymbol{m}| = 1, \quad (2)$$

where $M_s$ is the saturation magnetization, and $\boldsymbol{m}$ is the unit magnetization vector. The internal energy per unit volume of a magnetic material can be written in the form [29]:

$$E_{tot} = -\mu_0 \boldsymbol{H} \cdot \boldsymbol{M} + K(1 - (\boldsymbol{m} \cdot \boldsymbol{I})^2) + \frac{1}{2}\mu_0(\boldsymbol{N} \cdot \boldsymbol{M}) \cdot \boldsymbol{M}$$
$$- \mu_0 \boldsymbol{h_{dip}} \cdot \boldsymbol{M} + E_{me} + E_{el}, \quad (3)$$

where $\boldsymbol{H}$ is the external magnetic field, $K$ is the first-order uniaxial anisotropy, $\boldsymbol{I} = (cos\psi, sin\psi, 0)$ is the uniaxial anisotropy direction, and $\boldsymbol{N}$ is the tensor demagnetization factor. $\boldsymbol{h_{dip}}$ is the dipole-dipole interaction field resulting from the inhomogeneity distribution of magnetic moments in the space, and given by the magnetostatic Green's function in the Fourier space $G_k(z)$ [29–32],

$$\boldsymbol{h_{dip,k}} = \int_L G_k(z - z')\boldsymbol{m_k}(z')\, dz'. \quad (4)$$

$E_{me}$ and $E_{el}$ are the magneto-elastic coupling and the elastic energies, respectively, and given by [29]

$$E_{me} = B_1(\eta_{11}m_1^2 + \eta_{22}m_2^2 + \eta_{33}m_3^2)$$
$$+ B_2(\eta_{12}m_1 m_2 + \eta_{13}m_1 m_3 + \eta_{23}m_2 m_3), \quad (5)$$

$$E_{el} = \frac{1}{2}c_{ijkl}\eta_{ij}\eta_{kl} \quad i,j,k,l \in \{1,2,3\}. \quad (6)$$

Here, $B_1$ and $B_2$ are the magnetoelastic coupling coefficients, and $B_1 = B_2$ for isotropic amorphous films. $\eta_{ij}$ are the strain tensor, and $c_{ijkl}$ are the tensor of the elastic constants. In Eq. (6) and below, the repeating indices $(i, j, k, l)$ are assumed to be summed.

As we are interested in the acoustic wave propagation, we consider only small dynamic perturbations around an equilibrium orientation of magnetic moment for a given external magnetic field $H$,

$$\boldsymbol{m} = \boldsymbol{m_0} + \delta\boldsymbol{m}, \qquad (7)$$

with $\delta\boldsymbol{m} \ll \boldsymbol{m_0}$. The Landau-Lifshitz-Gilbert (LLG) equation in spherical coordinates is as follows:

$$\begin{pmatrix} \alpha & -sin\theta_0 \\ sin\theta_0 & \alpha sin^2\theta_0 \end{pmatrix} \begin{pmatrix} \frac{\partial \theta}{\partial t} \\ \frac{\partial \varphi}{\partial t} \end{pmatrix} = -\frac{\gamma}{\mu_0 M_s} \begin{pmatrix} \frac{\partial E_{tot}}{\partial \theta} \\ \frac{\partial E_{tot}}{\partial \varphi} \end{pmatrix} \qquad (8)$$

with the damping factor $\alpha$, and the gyromagnetic ratio $\gamma$. $(\theta_0, \varphi_0)$ define the orientation of the magnetic moment in the equilibrium state with the minimum of local energy. According to Stoner-Wohlfarth model [33], $(\theta_0, \varphi_0)$ can be calculated for any particular $\boldsymbol{H}$ by finding the zeros of $\partial E_{tot}/\partial \theta$ (or $\partial E_{tot}/\partial \varphi$) for which $\partial^2 E_{tot}/\partial \theta^2 > 0$ (or $\partial^2 E_{tot}/\partial \varphi^2 > 0$). The acoustic perturbation rotates the magnetization to a new $\theta$ and $\varphi$ orientation, where $\theta = \theta_0 + \delta\theta$ and $\varphi = \varphi_0 + \delta\varphi$. $\delta\theta$ and $\delta\varphi$ are the offset angles under the disturbance with $\delta\theta \ll \theta$、$\delta\varphi \ll \varphi$. Consider an infinite magnetic thin film of thickness $d$ with its normal parallel to the z-direction (Fig. 2), the dipolar field can be approximated as [31]

$$\boldsymbol{h_{dip}} = -M_s \left[ \left(1 - \frac{1-e^{-kd}}{kd}\right) \frac{\boldsymbol{k} \cdot \delta\boldsymbol{m}}{k^2} \boldsymbol{k} + \frac{1-e^{-kd}}{kd}(\vec{z} \cdot \delta\boldsymbol{m})\vec{z} \right], \qquad (9)$$

with

$$\delta\boldsymbol{m} = \begin{pmatrix} cos\theta_0 cos\varphi_0 & -sin\theta_0 sin\varphi_0 \\ cos\theta_0 sin\varphi_0 & sin\theta_0 cos\varphi_0 \\ -sin\theta_0 & 0 \end{pmatrix} \begin{pmatrix} \delta\theta \\ \delta\varphi \end{pmatrix}. \qquad (10)$$

Then, the energy is expanded around the equilibrium position:

$$E_{tot} = E_0 + \frac{E_{\theta\theta}}{2}\delta\theta^2 + \frac{E_{\varphi\varphi}}{2}\delta\varphi^2 + E_{\theta\varphi}\delta\theta\delta\varphi + E_{\theta\eta_{ij}}\delta\theta\delta\eta_{ij} + E_{\varphi\eta_{ij}}\delta\varphi\delta\eta_{ij} \quad (11)$$

$E_{\theta\theta}$, $E_{\varphi\varphi}$, $E_{\theta\varphi}$, $E_{\theta\eta_{ij}}$ and $E_{\varphi\eta_{ij}}$ are the second-order derivation of $E_{tot}$ to $\theta$, $\varphi$ and $\eta_{ij}$, respectively. Making a plane-wave ansatz for offset angles $\delta\theta = \vartheta\exp(i(\mathbf{kr} - \omega t))$ and $\delta\varphi = \Phi\exp(i(\mathbf{kr} - \omega t))$, Eq. (8) can be solved by $\delta\theta$ and $\delta\varphi$ and written as

$$\begin{pmatrix}\partial\theta\\\partial\varphi\end{pmatrix} = \frac{1}{D}\begin{pmatrix}\frac{i\omega\alpha}{\gamma}\sin^2\theta_0 - \frac{E_{\varphi\varphi}}{\mu_0 M_s} & \frac{i\omega}{\gamma}\sin\theta_0 + \frac{E_{\theta\varphi}}{\mu_0 M_s}\\-\frac{i\omega}{\gamma}\sin\theta_0 + \frac{E_{\theta\varphi}}{\mu_0 M_s} & \frac{i\omega\alpha}{\gamma} - \frac{E_{\theta\theta}}{\mu_0 M_s}\end{pmatrix}\begin{pmatrix}h_\theta\\h_\varphi\end{pmatrix}, \quad (12)$$

with

$$D = \left(\frac{i\omega\alpha}{\gamma}\sin^2\theta_0 - \frac{E_{\varphi\varphi}}{\mu_0 M_s}\right)\left(\frac{i\omega\alpha}{\gamma} - \frac{E_{\theta\theta}}{\mu_0 M_s}\right) - \left(\frac{\omega}{\gamma}\sin\theta_0\right)^2 - \frac{E_{\theta\varphi}^2}{(\mu_0 M_s)^2}. \quad (13)$$

As shown in Fig. 2, $h_\theta$ and $h_\varphi$ are the equivalent driving fields of the strain respectively along the $\theta$ and $\varphi$ directions

$$\begin{pmatrix}h_\theta\\h_\varphi\end{pmatrix} = \frac{1}{\mu_0 M_s}\begin{pmatrix}E_{\theta\eta_{ij}}\\E_{\varphi\eta_{ij}}\end{pmatrix}\eta_{ij}. \quad (14)$$

Now, we get the offset angles caused by acoustic wave in any initial condition. At this point, a stress-strain constitutive equation of the magnetoelastic material needs to be found. A First Piola-Kirchoff stress tensor can be defined as [34]

$$\sigma_{ij} = \frac{\partial E_{tot}}{\partial\eta_{ij}} = \frac{\partial E_{el}}{\partial\eta_{ij}} + \frac{\partial E_{me}}{\partial\eta_{ij}} = (c_{ijkl} + \Delta c_{ijkl})\eta_{kl}. \quad (15)$$

As can be seen, the extra magnetoelastic term, $E_{me}$, is the source of the *ΔE/ΔG* effect, with

$$\Delta c_{ijkl} = \frac{\partial^2 E_{me}}{\partial\eta_{ij}\partial\eta_{kl}}. \quad (16)$$

Equation (16) defines the relationship between the elastic tensor and the magnetoelastic energy term. The shear modulus G is equivalent to the $\Delta c_{1212}$ (abbreviated as $\Delta c_{66}$). If **m** lies within the film plane ($\theta_0 = 90°$), we find the relationship between $\Delta c_{66}$ of the magnetic film and the local internal energy

$$\Delta c_{66} = \frac{B_2^2 \cos^2 2\varphi_0}{\mu_0 M_s D}\left(\frac{i\omega\alpha}{\gamma} - \frac{E_{\theta\theta}}{\mu_0 M_s}\right). \quad (17)$$

In Eq. (17), terms above the quadratic are ignored in **m**. Three independent parameters, $\omega$, **H**, and $\psi$, contribute to the $\Delta G$ effect. In the frequency range of $\omega \ll \omega_s$ ($\omega_s$ represents the natural ferromagnetic resonance frequency), Eq. (17) can be further reduced to

$$\Delta c_{66}(H,\psi) = -\frac{B_2^2 \cos^2 2\varphi_0}{E_{\varphi\varphi}}, \quad (18)$$

with

$$E_{\varphi\varphi} = \mu_0 \mathbf{H} \cdot \mathbf{M} + 2K\cos 2(\varphi_0 - \psi) + E_{\varphi\varphi}^{me} + \mu_0 M_s^2\left(1 - \frac{1-e^{-kd}}{kd}\right)\sin^2\varphi_0, \quad (19)$$

which was reported in Ref. [27] except the last term. Later, we will discuss the variation trend of $\Delta c_{66}$ with respect to **H** and $\psi$ in details.

**B. FEM Analysis of the Magneto-acoustic Coupling**

To illustrate the coupling between the elastic wave and the magnetization dynamics in a multi-layer structure, a frequency domain FEM model is built using a commercial COMSOL Multiphysics simulation software, which includes the micromagnetics, piezoelectrics, elastic dynamics, and the feedback of magnetostrictive strain from the magnetic system to the elastic system in a fully coupled manner.

A ST-cut 90°x quartz is chosen as the piezoelectric substrate with Euler angles (0°, 132.75°, 90°). Al electrodes of 150 nm, a 0.8 μm $SiO_2$ waveguide layer and a 140 nm-thick FeCoSiB layer are subsequently constructed on the ST-cut 90°x quartz. The material constants of FeCoSiB thin film come from Ref. [22,35], whereas the rests are available in the software library. The length of the piezoelectric substrate is equal to the SAW wavelength $\lambda$ = 10 μm, and its thickness is set as 3$\lambda$. A perfectly matched layer and fixed constraints at the bottom prevent the interference of body reflected waves with SAWs. For simplicity, only one acoustic wavelength scale is simulated, and periodic boundary conditions are imposed on both sides of the sound propagation direction, which corresponds to extending the structure infinitely to both sides.

The behavior of ST-cut 90°x quartz follows the linear piezoelectric equation

$$\boldsymbol{\sigma} = \boldsymbol{C}^E \boldsymbol{\eta}^{el} - \mathbf{e}\boldsymbol{E},$$
$$\boldsymbol{D} = \boldsymbol{e}^T \boldsymbol{\eta}^{el} + \boldsymbol{\xi}\,\boldsymbol{E}, \tag{20}$$

where $\boldsymbol{D}$ and $\boldsymbol{E}$ are the electrical displacement and electric field, respectively. $\boldsymbol{C}^E$, $\boldsymbol{\xi}$ and $e$ are the elastic stiffness, permittivity tensor and piezoelectric coupling tensor, respectively. As mentioned in Section. II A, the damped spin behavior in a magnetic field can be described by the micromagnetic Eq. (8). Equation (21) donates the strain-displacement relationship for magnetoelastic materials, defining the bidirectional coupling between magnetoelasticity and elastic strain, given by

$$\boldsymbol{\eta}^{tot} = \boldsymbol{\eta}^{el} + \boldsymbol{\eta}^{me} = \frac{1}{2}[(\nabla \boldsymbol{u})^T + \nabla \boldsymbol{u}]. \tag{21}$$

The elastodynamic process describes the entire system using Newton's equations

$$-\rho \omega^2 \boldsymbol{u} = \nabla \cdot \boldsymbol{\sigma}, \tag{22}$$

where $\boldsymbol{\eta}^{me}$, $\rho$, $\boldsymbol{u}$ are the magnetostrictive strain tensor, the material density and the displacement field, respectively.

The theoretical formulas above are available in the corresponding solid mechanics and piezoelectrics modules, while the micromagnetic equations are developed utilizing weak form partial differential equation modules, which are coupled to the solid mechanics modules. This model allows us to analyze magneto-acoustic-electric coupling within the multilayered structure. Additionally, the resonance frequency, deformation and admittance changes of the structure caused by the magnetic field can be extracted from the eigenfrequency and frequency domain calculations.

### III. METHODS OF DEVCE FABRICATION AND MEASUREMENT

Magnetic SAW resonators were fabricated on ST-cut 90°X quartz substrates. Aluminum IDTs with a thickness of 150 nm were deposited by thermal evaporation, and then patterned by a photolithography lift-off process. All resonators consist of 100 pairs IDTs and two sets of 60 electrode reflectors on each end of the device with a wavelength ($\lambda$) of 10 μm and pitch of 200$\lambda$. A SiO$_2$ layer was deposited onto IDTs and thinned to a thickness of 800 nm by chemical mechanical polishing (CMP). A 140 nm-thick FeCoSiB film was then deposited on the SiO$_2$ layer by magnetron sputtering using a (Fe$_{90}$Co$_{10}$)$_{78}$Si$_{12}$B$_{10}$ target. An in-situ magnetic field ~200 Oe is applied during sputtering to induce in-plane uniaxial anisotropy. Six samples with different angles ($\psi$ = 0°, 8°, 22.5°, 45°, 67.5°, and 90°, respectively) between the easy axis (EA) and the SAW propagation direction have been fabricated, denoted as D1 to D6 devices.

Fig.3(a) shows the in-plane magnetization versus external magnetic field (*M–H*) curves for the FeCoSiB film measured by a vibrating sample magnetometer (*VSM, BHV-525, Japan*). The sample shows a clear easy axis with an in-plane saturation field $H_k$ of ~20 Oe and a low coercive field $H_c$ of 1.7 Oe. To obtain the high frequency properties of film, we have further measured the permeability spectra using a shorted microstrip transmission-line perturbation method [36], shown in Fig. 3(b). The measured natural ferromagnetic resonance is about 1.7 GHz, much higher than the operating frequency of SAW resonators. Fitting the imaginary permeability spectrum yields an effective damping factor of $\alpha = 0.009$, which ensures the low FMR loss at the operating frequency.

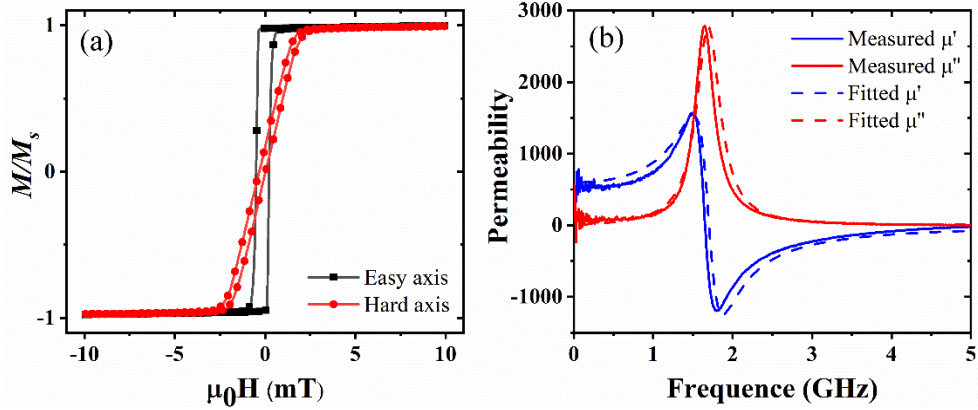

FIG. 3. **(a)** M-H curves of the FeCoSiB thin film along the easy (black) and hard axis (red), and **(b)** measured and fitted permeability spectra from 10 MHz to 5 GHz.

The frequency response of the SAW resonators upon applying an external magnetic field was measured by tracing the $S_{11}$ scattering parameter using an Agilent network analyzer (N5230A). A rotatable Helmholtz coil driven by a current source (ITECH-6502A) was used to provide uniform in-plane magnetic field. Continuous magnetic field sweep ranging between $\pm 80$ Oe is sufficient to magnetize the FeCoSiB film to saturation in any measuring direction. A Gauss meter (Lake Shore 425) was used to calibrate magnetic field. All experiments were carried out at room temperature.

## IV. RESULTS AND DISCUSSION

## A. Field dependent magneto-acoustic coupling

Fig. 4 show that the $S_{11}$ spectrum of a bare SAW resonator with no magnetic overlayer exhibits a strong resonance peak at ~477 MHz. The presence of a 140 nm-thick FeCoSiB layer with $\psi = 0°$ (D1) significantly down shifts the resonance frequency ($f_r$) to ~447 MHz, due to the lower shear wave velocity of FeCoSiB (2029 m/s, calculate by $\sqrt{\frac{c_{44}}{\rho}}$) compare with those of $SiO_2$ and ST-cut 90°X quartz, 3748 and 5047m/s, respectively. Meanwhile, the amplitude of resonance peak decays from 8 dB to 0.9 dB. It can be attributed to the large extra capacitance between the magnetic film and the IDT, together with the power absorption by the magneto-acoustic coupling [37,38]. FEM simulations show that the displacements of both devices at the resonance frequency can be assigned to the Love mode resonance. The inset in Fig. 4 also shows the photo of a packaged SAW resonator.

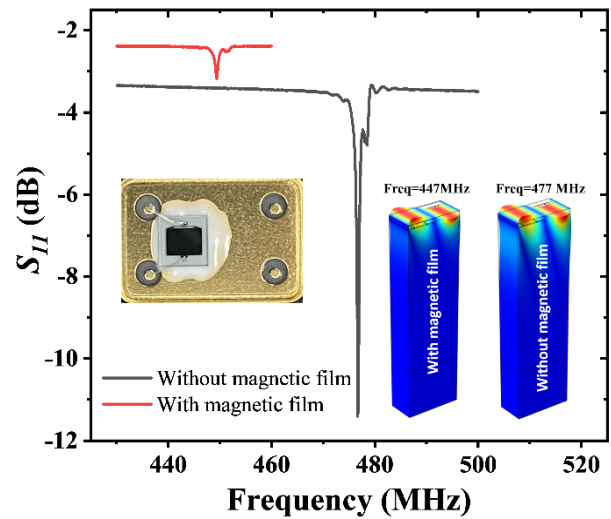

FIG. 4. Measured $S_{11}$ as a function of frequency from SAW resonators with and without FeCoSiB film. Inset shows the photo of a packaged device. The displacements at resonance frequencies are obtained through COMSOL simulations.

The magneto-acoustic responses of D1 and D6 devices are then measured upon applying magnetic field along the hard axis, as schematically illustrated in the inset of Fig. 5(a) and 5(d). A test cycle consists of three steps: initialization (magnetic field from 0 to 80 Oe), backward sweep (80 Oe to -80 Oe), and forward sweep (-80 Oe to 80 Oe). Despite a small offset caused by hysteresis, the frequency responses of the three sweeps almost overlaps, so only the forward sweep is plotted here. Fig. 5(a) and Fig. 5(d) plot the resonance frequency shift of D1 and D6, respectively, which clearly show the difference between the two devices. A strong ΔG effect is observed for D1 at low sweeping fields from -15 to +15

Oe, corresponding to a maximum frequency shift ($\Delta f_{max}$) of 3.9 MHz. The highest *RFS* (or d*f*/d*H*) is about 634 kHz/Oe under an external field of 3.3 Oe. However, the ΔG effect is negligible for D6 at low sweeping fields, and only becomes much stronger at high fields above |15 Oe|. $\Delta f_{max}$ is about 6.8 MHz and the highest *RFS* reaches 775 kHz/Oe under a field of -16.6 Oe. To our best knowledge, this sets a record for Love-mode SAW magnetic field sensors. The frequency response of former has been reported in previous studies [11], and was interpreted from a magnetic mean field model, i.e. the angular dispersion of magnetic moments, arising from stray fields, local stresses, or an intrinsic distribution of the easy axis. However, it cannot explain the frequency response or the ΔG effect observed from the D6 device.

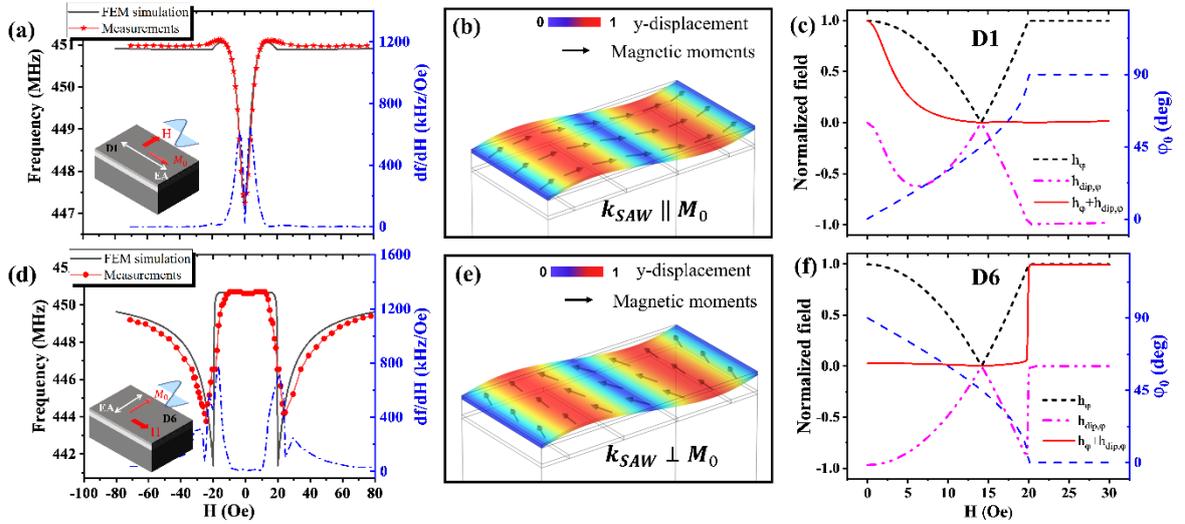

FIG. 5. Results of D1 and D6 devices along the hard axis. **(a)** and **(d)** Measured and simulated frequency response as a function of magnetic field. Insets illustrate the measurement set-up where $M_0$ is parallel to $k_{SAW}$ for D1, but is perpendicular to $k_{SAW}$ for D6, and the sweep field is perpendicular to $M_0$. **(b)** and **(e)** Simulated displacements and magnetic moment distributions at resonance frequency. **(c)** and **(f)** Calculated magnetoelastic driven field $h_\varphi$, in-plane dipole field $h_{dip,\varphi}$ and the effective perturbation field $h_\varphi + h_{dip,\varphi}$, all of which are normalized. The vertical axis on the right shows the variation of magnetic moment orientation as a function of applied magnetic field.

To understand the nature of the resonance frequency response, FEM simulations were carried out using the magneto-acoustic model in Section. II. As shown in Fig. 5(b) and 5(e), the displacements and angular distributions of the magnetic moment in the FeCoSiB film are extracted in the case of $\boldsymbol{k_{SAW}} \| \boldsymbol{M_0}$ and $\boldsymbol{k_{SAW}} \perp \boldsymbol{M_0}$. Due to the smaller wavelength of the magneto-elastic wave compared to the film size, magnetic moments are non-uniformly perturbed, leading to a non-zero phase relationship between adjacent magnetic moments. In contrast to Fig. 5(b), a non-zero phase distribution of the magnetic moment perpendicular to $\boldsymbol{k_{SAW}}$ in Fig. 5(e) yields a dipole field $h_{dip}$. Using Eq. (9), (10) and (12), for $\omega \ll \omega_s$, the in-plane component of dipole field parallel to $h_\varphi$ can be expressed as

$$h_{dip,\varphi} = -M_s \frac{sin^2\varphi_0}{E_{\varphi\varphi}}\left(1 - \frac{1-e^{-kd}}{kd}\right)h_\varphi, \tag{23}$$

Therefore, Eq. (18) can be rewritten as

$$\Delta c_{66} = -\frac{B_2 cos2\varphi_0}{\mu_0 \boldsymbol{H} \cdot \boldsymbol{M} + E_{\varphi\varphi}^{me} + 2Kcos2(\varphi_0 - \psi)} \frac{h_\varphi + h_{dip,\varphi}}{\eta_{12}}, \tag{24}$$

where $h_\varphi + h_{dip,\varphi}$ is defined as the effective perturbation field. Normalized $h_\varphi$, $h_{dip,\varphi}$ and $h_\varphi + h_{dip,\varphi}$ are calculated as a function of magnetic field $\boldsymbol{H}$ using Eq. (14) and (23), and plotted in Fig. 5(c) and 5(f), assuming a constant $\eta_{12}$. As far as $\boldsymbol{H}$ is applied along the hard axis, there is a consistent relationship between $h_\varphi$ and $\boldsymbol{H}$ for both D1 and D6 devices, but $h_{dip,\varphi}$ heavily depends on the angle ($\varphi_0$) between $\boldsymbol{M}_0$ and $\boldsymbol{k_{SAW}}$. When $\varphi_0$ is close to 90° (high field region in Fig. 5(c) and low field region in Fig. 5(f)), $|h_{dip,\varphi}|$ is significantly enhanced, but when $\varphi_0$ is close to 0°, $|h_{dip,\varphi}|$ reduces to zero. In Fig. 5(f), the sum of $h_\varphi$ and $h_{dip,\varphi}$ is quite small in the range of -20 to +20 Oe, leading to a suppressed $\Delta G$ effect according to Eq. (24). As seen in Fig. 5(a) and 5(d), our FEM simulation results (black solid line) are in general accordance with the experiments.

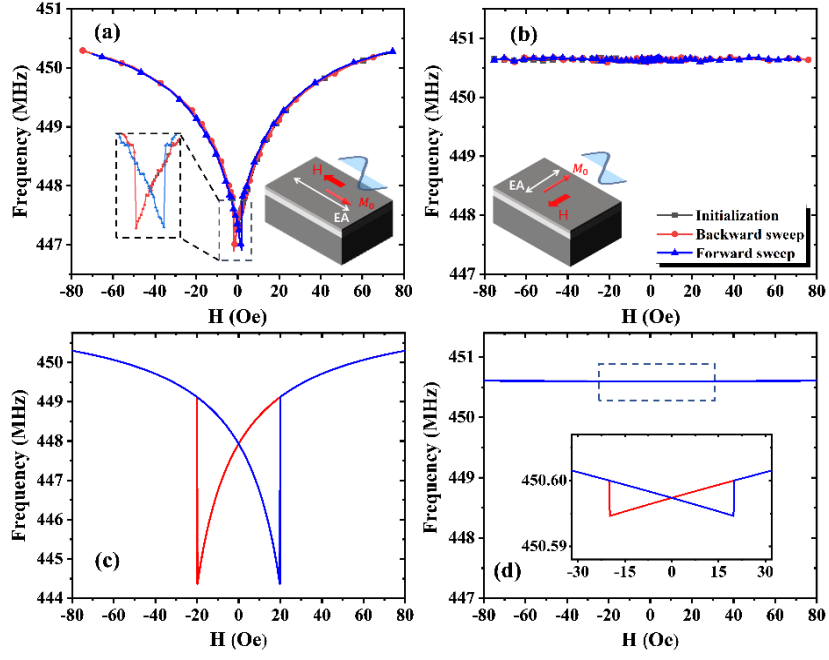

FIG. 6. Measured, (a) & (b), and simulated, (c) and (d), frequency responses of D1 and D6 devices, respectively, upon applying field along the easy axis. Inset shows the amplified frequency response.

The measured frequency responses of the D1 and D6 devices are also significantly different from each other upon applying field along the easy axis. As shown in Fig. 6(a), during both forward and backward sweeps, D1 exhibits almost the similar magnitude of $\Delta f_{max}$ as that in Fig. 5(a), and $f_r$ jumps at small magnetic fields (±1.94 Oe). Both of them can be explained by the term $\mu_0 \boldsymbol{H} \cdot \boldsymbol{M}$ in Eq. (24), which is enhanced by a parallel applied field $\boldsymbol{H}$ and reduced by an antiparallel one since $\varphi_0 = \psi$. Meanwhile, magnetic moments flip 180° upon applying an antiparallel field, causing $\Delta c_{66}$ to jump near the coercive fields (see Fig. 3a). For D6 with $\boldsymbol{k_{SAW}} \perp \boldsymbol{M_0}$, the initial magnetic moment orientation $\varphi_0$ is 90° and doesn't change with the applied field, which corresponds to a large $h_{dip,\varphi}$. This makes the second term ($h_\varphi + h_{dip,\varphi}$) in Eq. (24) very small, and thus the ΔG effect is attenuated. Fig. 6(c) and 6(d) plot the simulated results of D1 and D6 upon applying magnetic field along the EA. According to Fig. 6(d), the amplitude of frequency response of D6 is around 2 orders of magnitudes weaker than that of D1, and thus is nearly buried by noise. Therefore, almost no frequency response can be observed from D6 (Fig. 6(b)).

In this section, a dipole field is used to explain the different frequency responses of D1 and D6, which are in good agreement with the experimental results. The deviation between the measured and simulated ones, especially between Fig. 6(a) and Fig. 6(c), may be attributed to dispersed magnetic moments or magnetic domains, which are not considered in the Stoner-Wohlfarth model. Fig. 5(d) together with Fig. 6(b) demonstrate that D6 is a true vector magnetic field sensor. However, notice that a bias field of ~16.6 Oe is still required to obtain a high *RFS*.

## B. Angle dependent magneto-acoustic coupling

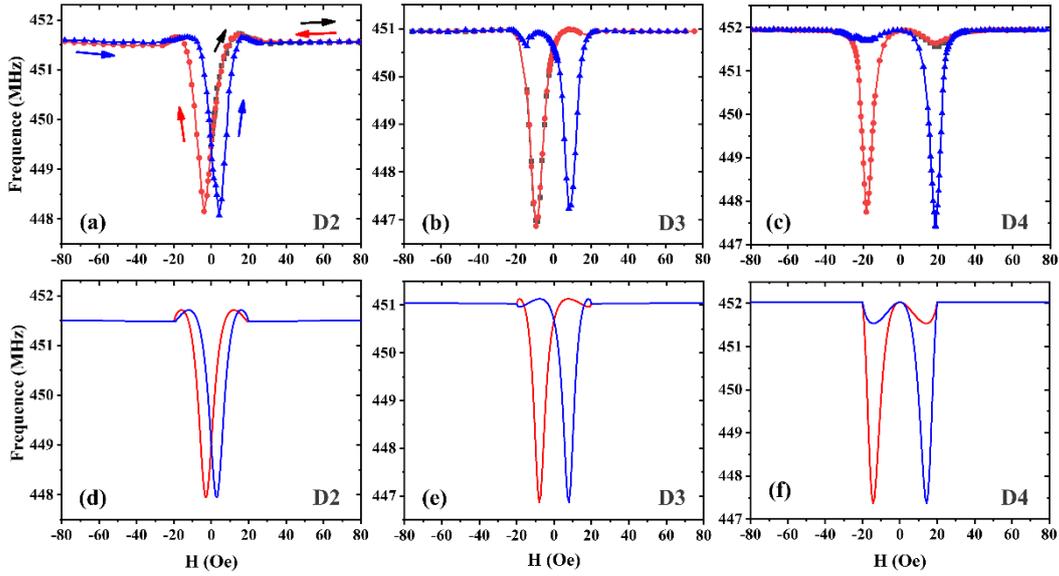

FIG. 7. Measured (top panel) and simulated (bottom panel) frequency responses of D2 **(a** and **d)**, D3 **(b** and **e)** and D4 **(c** and **f)** devices. Magnetic field is applied along hard axis. The test cycle includes initialization (black line), backward sweep (red line) and forward sweep (blue line).

Let's now turn to the resonance frequency response of D2, D3, and D4 devices ($\psi$ = 8°, 22.5° and 45°). The measured frequency and simulated responses upon applying a magnetic field along the hard axis are plotted on the top and bottom panels of Fig. 7. Compared to the D1 device, D2, D3 and D4 show butterfly-like frequency response curves, and the minimum resonance frequency ($f_{min}$) shifts to the right (or left) side during the forward (or backward) sweep. The higher $\psi$ angle is, the larger $f_{min}$ offset is seen in Fig. 7 (a-c). The steep frequency response at zero field ensures the high sensitivity of these devices even without a bias field.

In Equation (24), $\Delta c_{66}$ exhibits a pronounced $\varphi_0$ dependence, which is controlled by both $H$ and $\psi$. In other words, specific magnetization orientation $\varphi_0$ can be referred to different combinations of ($H$, $\psi$). As shown in Fig. 5a, the magnetic moment is rotated a special $\varphi_0$ angle (about 9.5°) by applying a 3.3 Oe magnetic field to maximize $RFS$. Likewise, this angle can also be achieved by directly setting $\psi = 9.5°$ during magnetic film deposition. Fig. 8 plots the calculated $\Delta c_{66}$ with magnetic field along the hard axis for $\psi = 0°$ and $\psi = 9.5°$, respectively. The change of $\psi$ from 0° to 9.5° shifts the entire curve from right to left, which is almost equivalent to a bias field of 3.3 Oe, although the magnitude of $\Delta c_{66}$ is slightly different. As a result, a non-zero $RFS$ can be obtained at zero bias field.

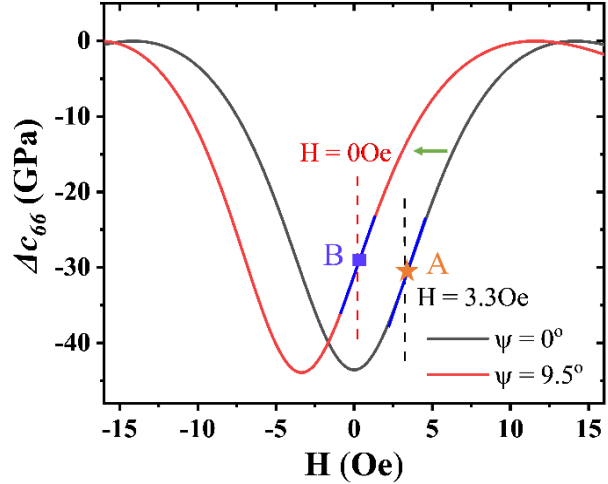

FIG .8. Calculated variation of $\Delta c_{66}$ with magnetic field along the hard axis when $\psi = 0°$ and $\psi = 9.5°$, respectively. At point A, a magnetic field of 3.3 Oe rotates magnetic moment orientation $\varphi_0$ from 0° to 9.5°. At point B, magnetic moments are aligned at $\varphi_0 = 9.5°$ under zero bias field by setting $\psi = 9.5°$.

One obvious strategy to improve the self-biased $RFS$ is to select an appropriate $\psi$ that maximizes $\left.\frac{\partial f}{\partial H}\right|_{H=0}$. Next, we calculated the relationship between $\left.\frac{\partial f}{\partial H}\right|_{H=0}$ and $\psi$ in the range of $\psi \in [0°, 90°]$. As shown in Fig. 9, self-biased *RFS* values can be obtained at almost any $\psi$ angles except 0°, 45°, and 90°, and the highest $\left.\frac{\partial f}{\partial H}\right|_{H=0}$ appears when $\psi \approx 7°$. As mentioned above, a large $\boldsymbol{h_{dip,\varphi}}$ suppresses the *ΔG* effect when $\psi$ ($\psi = \varphi_0$ in this case) is close to 90°. This will deteriorate *RFS*, thus making it unsuitable to select $\psi$ in the interval [45°, 90°]. We have measured the *RFS* values of D1-D6 devices ($\psi$ = 0°, 8°, 22.5°, 45°, 67.5°, and 90°) at zero field and marked in Fig. 9 as well. Again, the measured results are in good agreement with the simulated ones. It should be emphasized that the optimal $\psi$ angle with the maximum self-biased *RFS* is strongly affected by $\boldsymbol{h_{dip}}$. For different operating frequencies, magnetostrictive materials, layer thicknesses, or structures of devices, the optimal $\psi$ angle must be recalculated. For example, a 20 nm-thick FeCoSiB film has an optimal $\psi$ angle of 15° instead of 7° according to our simulation.

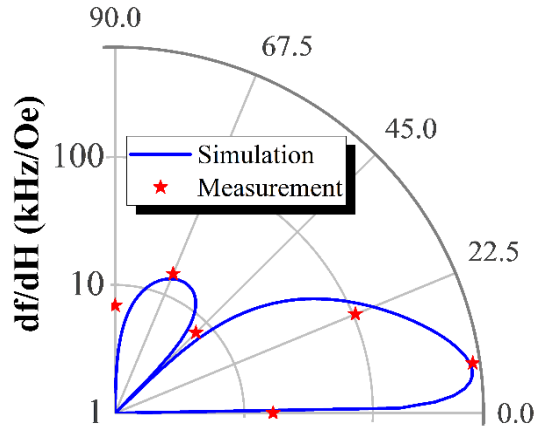

FIG. 9. Simulated and measured $\psi$ angle dependent *RFS* at zero bias field. Measurement results are extracted from D1-D6 devices with $\psi$ = 0°, 8°, 22.5°, 45°, 67.5°, and 90°.

Finally, the magnetic field sensitivity of D2 was measured along different directions. Fig. 10 plots the calculated d*f*/d*H* and experimental results at zero magnetic field. An '8'-like shape is clearly seen. The angles reference to the maximum and minimum *RFS* locate at ~94° and 4°, respectively, slightly off HA and EA. The measured highest *RFS* at zero field is about 630.4 kHz/Oe. It demonstrates our sensor's ability to identify vector magnetic fields. For example, arbitrary magnetic field direction can be determined by comparing the frequency differences between three vertically placed sensors, which provides a solid foundation for future array applications of SAW magnetic field sensors.

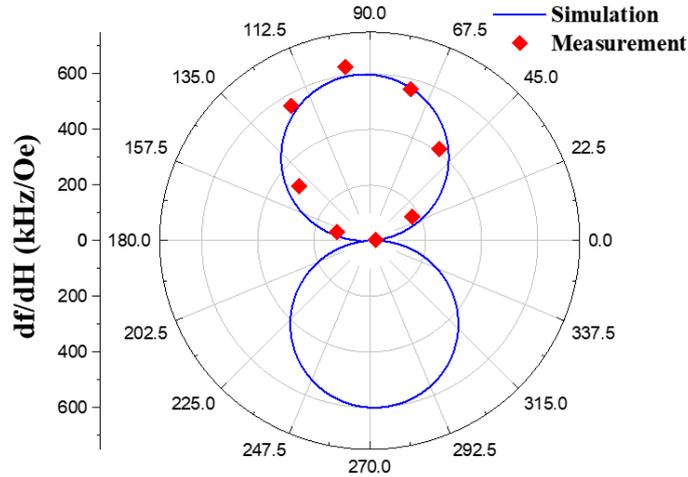

FIG. 10. Simulated and measured *RFS* results of D2 device at zero magnetic field. Experimental results are obtained by applying external magnetic field along different directions.

## V. Conclusion

In summary, a dynamic magnetoelastic model of a thin magnetostrictive film on a piezoelectric substrate has been established. The magneto-acoustic coupling between them was evaluated by measuring the resonance frequency shift of Love-mode SAW sensors. In comparison with traditional models, our dynamic magnetoelastic model considers the importance of the dipole-dipole interaction, and can explain the distinct frequency responses of SAW sensors with different $\psi$ angles between the acoustic wave vector and the in-plane induced anisotropy. Regulation of the dipole field have been achieved by setting an optimized $\psi$ angle during magnetic film deposition, which not only boosts the highest sensitivity of SAW magnetic sensors, but also yields a strong *RFS* of 630.4 kHz/Oe at zero bias field. This approach enables a simple implementation of self-biased SAW magnetic field sensors. Furthermore, the FEM simulation model developed in this work can be used for future investigations on the magneto-acoustic-electric coupling of other types of SAW magnetic field sensors or NEMS magnetoelectric antennas [39].

# Acknowledgement

This work is supported by the National Science Foundation of China (Grant No. 61871081) and Sichuan Science & Technology Support Program under Grant No. 2022GZ0267.

---

[1]   B. Gojdka, R. Jahns, K. Meurisch, H. Greve, R. Adelung, E. Quandt, R. Knöchel, and F. Faupel, Fully integrable magnetic field sensor based on delta-E effect, Appl. Phys. Lett. **99**, 223502 (2011).

[2]   A. V. Turutin, J. V. Vidal, I. V. Kubasov, A. M. Kislyuk, M. D. Malinkovich, Y. N. Parkhomenko, S. P. Kobeleva, O. V. Pakhomov, A. L. Kholkin, and N. A. Sobolev, Magnetoelectric metglas/bidomain y + 140°-Cut lithium niobate composite for sensing fT magnetic fields, Appl. Phys. Lett. **112**, 262906 (2018).

[3]   B. Spetzler, C. Bald, P. Durdaut, J. Reermann, C. Kirchhof, A. Teplyuk, D. Meyners, E. Quandt, M. Höft, G. Schmidt and F. Faupel, Exchange biased delta-E effect enables the detection of low frequency pT magnetic fields with simultaneous localization, Sci. Rep. **11**, 5269 (2021).

[4]   P. Smole, W. Ruile, C. Korden, A. Ludwig, E. Quandt, S. Krassnitzer, and P. Pongratz, Magnetically tunable SAW-resonator, in Proceedings of the Annual IEEE International Frequency Control Symposium (2003), pp. 903–906.

[5]   X. Liu, B. Tong, J. Ou-Yang, X. Yang, S. Chen, Y. Zhang, and B. Zhu, Self-biased vector magnetic sensor based on a Love-type surface acoustic wave resonator, Appl. Phys. Lett. **113**, 082402 (2018).

[6]   X. Liu, J. Ou-Yang, B. Tong, S. Chen, Y. Zhang, B. Zhu, and X. Yang, Influence of the delta-E effect on a surface acoustic wave resonator, Appl. Phys. Lett. **114**, 062903 (2019).

[7]   H. Mishra, J. Streque, M. Hehn, P. Mengue, H. M'Jahed, D. Lacour, K. Dumesnil, S. Petit-Watelot, S. Zhgoon, V. Polewczyk, A. Mazzamurro, A. Talbi, S. Hage-Ali and O. Elmazria, Temperature compensated magnetic field sensor based on love waves, Smart Mater. Struct. **29**, 045036 (2020).

[8]   H. Mishra, M. Hehn, S. Hage-Ali, S. Petit-Watelot, P. W. Mengue, S. Zghoon, H. M'Jahed, D. Lacour, and O. Elmazria, Microstructured Multilayered Surface-Acoustic-Wave Device for Multifunctional Sensing, Phys. Rev. Appl. **14**, 014053 (2020).

[9]   W. Li, P. Dhagat, and A. Jander, Surface acoustic wave magnetic sensor using Galfenol thin film, IEEE Trans. Magn. **48**, 4100 (2012).

[10]  H. Zhou, A. Talbi, N. Tiercelin, and O. Bou Matar, Multilayer magnetostrictive structure based surface acoustic wave devices, Appl. Phys. Lett. **104**, 114101 (2014).

[11]  A. Kittmann, P. Durdaut, S. Zabel, J. Reermann, J. Schmalz, B. Spetzler, D. Meyners, N. X. Sun, J.

McCord, M. Gerken, G. Schmidt, M. Höft, R. Knöchel, F. Faupel and E. Quandt, Wide band low noise love wave magnetic field sensor system, Sci. Rep. **8**, 278 (2018).

[12] A. Mazzamurro, Y. Dusch, P. Pernod, O. Bou Matar, A. Addad, A. Talbi, and N. Tiercelin, Giant Magnetoelastic Coupling in a Love Acoustic Waveguide Based on TbCo$_2$/FeCo Nanostructured Film on ST-Cut Quartz, Phys. Rev. Appl. **13**, 044001 (2020).

[13] J. Schmalz, A. Kittmann, P. Durdaut, B. Spetzler, F. Faupel, M. Höft, E. Quandt, and M. Gerken, Multi-Mode Love-Wave SAW Magnetic-Field Sensors, Sensors. **20**, 3421 (2020).

[14] M. Elhosnia, O. Elmazriaa, S. Petit-Watelota, L. Bouvota, S. Zhgoonb, A. Talbic, M. Hehna, K. A. Aissaa, S. Hage-Alia, D. Lacoura, F. Sarrya, and O. Boumatar, Magnetic field SAW sensors based on magnetostrictive-piezoelectric layered structures: FEM modeling and experimental validation, Sensors and Actuators, A: Physical **240**, 41 (2016).

[15] V. PolewczykV, K. Dumesnil, D. Lacour, M. Moutaouekkil, H. Mjahed, N. Tiercelin, S. P. Watelot, H. Mishra, Y. Dusch, S. Hage-Ali, O. Elmazria, F. Montaigne, A. Talbi, O. Bou Matar, and M. Hehn, Unipolar and Bipolar High-Magnetic-Field Sensors Based on Surface Acoustic Wave Resonators, Phys. Rev. Appl. **8**, 024001 (2017).

[16] T. Nan, Y. Hui, M. Rinaldi, and N. X. Sun, Self-Biased 215MHz Magnetoelectric NEMS Resonator for Ultra-Sensitive DC Magnetic Field Detection, Sci. Rep. **3**, 02115 (2013).

[17] M. Li, A. Matyushov, C. Dong, H. Chen, H. Lin, T. Nan, Z. Qian, M. Rinaldi, Y. Lin, and N. X. Sun, Ultra-sensitive NEMS magnetoelectric sensor for picotesla DC magnetic field detection, Appl. Phys. Lett. **110**, 143510 (2017).

[18] S. Pawar, J. Singh, and D. Kaur, Magnetic Field Tunable Ferromagnetic Shape Memory Alloy-Based Piezo-Resonator, IEEE Electron Device Lett. **41**, 280 (2020).

[19] J. Singh, A. Kumar, and M. Kumar, Highly Tunable Film Bulk Acoustic Wave Resonator Based on Pt/ZnO/Fe$_{65}$Co$_{35}$ Thin Films, IEEE Trans. Ultrason. Ferroelectr. Control **67**, 2130 (2020).

[20] A. Ludwig and E. Quandt, Optimization of the ΔE Effect in thin films and multilayers by magnetic field annealing, IEEE Trans. Magn. **38**, 2829 (2001).

[21] A. Piorra, R. Jahns, I. Teliban, J. L. Gugat, M. Gerken, R. Knöchel, and E. Quandt, Magnetoelectric thin film composites with interdigital electrodes, Appl. Phys. Lett. **103**, 032902 (2013).

[22] E. Lage, C. Kirchhof, V. Hrkac, L. Kienle, R. Jahns, R. Knöchel, E. Quandt, and D. Meyners, Exchange biasing of magnetoelectric composites, Nat. Mater. **11**, 523 (2012).

[23] J. D. Livingston, Magnetomechanical Properties of Amorphous Metals, Phys. Status Solidi (a) **70**, 591 (1982).

[24] P. T. Squire, Phenomenological model for magnetization, magnetostriction and ΔE Effect in field-annealed amorphous ribbons, J. Magn. Magn. Mater. **87**, 299 (1990).

[25] P. T. Squire, Domain model for magnetoelastic behaviour of uniaxial ferromagnets, J. Magn. Magn. Mater. **140**, 1829 (1995).

[26] Z. Sárközi, K. Mackay, and J. C. Peuzin, Elastic properties of magnetostrictive thin films using bending and torsion resonances of a bimorph, J. Appl. Phys. **88**, 5827 (2000).

[27] O. Bou Matar, J. F. Robillard, J. O. Vasseur, A. C. Hladky-Hennion, P. A. Deymier, P. Pernod, and V. Preobrazhensky, Band gap tunability of magneto-elastic phononic crystal, J. Appl. Phys. **111**, 054901 (2012).

[28] L. Huang, Q. Lyu, D. Wen, Z. Zhong, H. Zhang, and F. Bai, Theoretical investigation of magnetoelectric surface acoustic wave characteristics of ZnO/Metglas layered composite, AIP Adv. **6**, 015103 (2016).

[29] A. G. Gurevich and G. A. Melkov, Magnetization Oscillations and Waves (2020).

[30] K. J. Harte, Theory of Magnetization Ripple in Ferromagnetic Films, J. Appl. Phys. **39**, 1503 (1968).

[31] B. A. Kalinikos and A. N. Slavin, Theory of dipole-exchange spin wave spectrum for ferromagnetic films with mixed exchange boundary conditions, J. Phys. C: Solid State Phys. **19**, 7013 (1986).

[32] A. Hernández-Mínguez, F. Macià, J. M. Hernàndez, J. Herfort, and P. v. Santos, Large Nonreciprocal Propagation of Surface Acoustic Waves in Epitaxial Ferromagnetic/Semiconductor Hybrid Structures, Phys. Rev. Appl. **13**, 044018 (2020).

[33] C. Tannous and J. Gieraltowski, The Stoner-Wohlfarth model of ferromagnetism, Eur. J. Phys. **29**, 475 (2008).

[34] H. R. Hamidzadeh, L. Dai, and R. N. Jazar, Wave Propagation in Solid and Porous Half-Space Media, Vol. 9781461492696 (2014).

[35] M. Jovičević Klug, L. Thormählen, V. Röbisch, S. D. Toxværd, M. Höft, R. Knöchel, E. Quandt, D. Meyners, and J. McCord, Antiparallel exchange biased multilayers for low magnetic noise magnetic field sensors, Appl. Phys. Lett. **114**, 192410 (2019).

[36] Y. Liu, L. Chen, C. Y. Tan, H. J. Liu, and C. K. Ong, Broadband complex permeability characterization of magnetic thin films using shorted microstrip transmission-Line perturbation, Rev. Sci. Instrum. **76**, (2005).

[37] L. Dreher, M. Weiler, M. Pernpeintner, H. Huebl, R. Gross, M. S. Brandt, and S. T. B. Goennenwein, Surface acoustic wave driven ferromagnetic resonance in nickel Thin Films: Theory and experiment, Phys. Rev. B **86**, 134415 (2012).

[38] M. Küß, M. Heigl, L. Flacke, A. Hefele, A. Hörner, M. Weiler, M. Albrecht, and A. Wixforth, Symmetry of the Magnetoelastic Interaction of Rayleigh and Shear Horizontal Magnetoacoustic Waves in Nickel Thin Films on LiTaO$_3$, Phys. Rev. Appl. **15**, 034046 (2021).

[39] T. Nan, H. Lin, Y. Gao, A. Matyushov, G. Yu, H. Chen, N. Sun, S Wei, Z. Wang, M. Li, X. Wang, A. Belkessam, R. Guo, B. Chen, J. Zhou, Z. Qian, Y. Hui, M. Rinaldi, M. E. McConney, B. M. Howe, Z. Hu, J. G. Jones, G. J. Brown, and N. X. Sun, Acoustically actuated ultra-compact NEMS magnetoelectric antennas, Nat. Commun. **8**, 296 (2017).